# Dust grain shattering in protoplanetary discs: collisional fragmentation or rotational disruption?

Stéphane Michoulier ★ and Jean-François Gonzalez

*Univ Lyon, Univ Claude Bernard Lyon 1, ENS de Lyon, CNRS, Centre de Recherche Astrophysique de Lyon UMR5574, F-69230, Saint-Genis-Laval, France*

## ABSTRACT

In protoplanetary discs, the coagulation of dust grains into large aggregates still remains poorly understood. Grain porosity appears to be a promising solution to allow the grains to survive and form planetesimals. Furthermore, grain shattering has generally been considered to come only from collisional fragmentation; however, a new process was recently introduced, rotational disruption. We wrote a one-dimensional code that models the growth and porosity evolution of grains as they drift to study their final outcome when the two shattering processes are included. When simulating the evolution of grains in a disc model that reproduces observations, we find that rotational disruption is not negligible compared to the fragmentation and radial drift. Disruption becomes dominant when the turbulence parameter $\alpha \lesssim 5 \times 10^{-4}$, if the radial drift is slow enough. We show that the importance of disruption in the growth history of grains strongly depends on their tensile strength.

## 1 INTRODUCTION

In the theory of planet formation, the growth of sub-μm to mm dust aggregates in protoplanetary discs into planetesimals is hampered by theoretical problems commonly known as the radial-drift barrier (Weidenschilling 1977) and the fragmentation barrier (Weidenschilling & Cuzzi 1993; Stepinski & Valageas 1996; Dominik & Tielens 1997; Blum & Wurm 2008). In the former, gas-drag induced inwards radial motion of the disc depletes the solids in the disc before they can grow to large sizes while in the latter, large collision velocities between aggregates lead to shattering rather than coagulation. Both barriers prevent dust grains to ultimately create planets. Numerous solutions have been proposed to trap dust in pressure maxima and form planetesimals, such as vortices (Barge & Sommeria 1995; Meheut et al. 2012; Loren-Aguilar & Bate 2015), snow lines (Kretke & Lin 2007; Brauer, Henning & Dullemond 2008b; Drazkowska, Windmark & Dullemond 2014; Vericel & Gonzalez 2020), or self-induced dust traps (Gonzalez, Laibe & Maddison 2017; Vericel & Gonzalez 2020; Vericel et al. 2021). Other processes based on instabilities are also studied, like streaming instabilities (Youdin & Goodman 2005; Youdin & Johansen 2007; Schäfer, Yang & Johansen 2017; Yang, Johansen & Carrera 2017; Auffinger & Laibe 2018; Li, Youdin & Simon 2019) or coagulation instabilities (Tominaga, Inutsuka & Kobayashi 2021). Another solution to overcome these barriers is to consider intrinsic dust properties, namely grain porosity. Grains are often considered compact for simplicity (Brauer, Dullemond & Henning 2008a; Drazkowska et al. 2014; Gonzalez et al. 2015; Vericel et al. 2021); however, grain properties like porosity could play a major role in their evolution (Ormel, Spaans & Tielens 2007; Suyama, Wada & Tanaka 2008). For a given mass, fluffy aggregates have a larger collisional cross-section, allowing them to grow faster and decouple rapidly at larger sizes, ensuring their survival in the disc. Garcia (2018) also showed that porous grains are less sensitive to fragmentation than compact grains and lead to planetesimal formation via coagulation.

Recently, rotational disruption of porous dust grains was proposed as another possible barrier and has been investigated by Tatsuuma & Kataoka (2021) in the framework of protoplanetary discs. They found that grains can be disrupted by the gas-flow torque when aggregates tend to be highly porous, before they can decouple from the gas, when very compact grains are not. In their study, grains evolve at fixed locations in an inviscid disc, considering radiative and gas flow torque, using the model initially developed by Okuzumi, Tanaka & Sakagami (2009), Okuzumi et al. (2012), and Kataoka et al. (2013).

In this paper, we study the behaviour of dust grains in different models of discs to understand in which case each shattering process, fragmentation due to collision between grains, or rotational disruption dominates, when they are allowed to move in the disc. We first introduce our code and the models we use for radial motion, growth, porosity evolution, fragmentation, and rotational disruption in Section 2. We then analyse our results based on several simulations to understand in which circumstances rotational disruption plays a role and influences the dust behaviour depending on disc models, monomer size, and tensile strength formulations derived by Tatsuuma, Kataoka & Tanaka (2019) and Kimura et al. (2020) in Section 3. Finally, we discuss our results and the limitations of our code in Section 4, and conclude in Section 5.

## 2 METHODS

### 2.1 The one-dimensional code

We present PAMDEAS (Porous Aggregate Model and Dust Evolution in protoplAnetary discS), a one-dimensional (1D) code designed to study the physics of porous grains in protoplanetary discs similar to PACED (Garcia & Gonzalez 2020). Our code includes several physical processes such as radial drift, grain growth, porosity evolution, and fragmentation, similarly to PACED. We added a better prescription for gas drag which includes all Stokes regimes, a correct formula of the orbital velocity of dust consistent with the radial drift velocity, and we implemented rotational disruption, as described in the following subsections.

★ E-mail: stephane.michoulier@univ-lyon1.fr

To model protoplanetary discs, we adopt the commonly used power-law formulation, for which two indices $p$ and $q$ are defined to express the gas surface density profile $\Sigma_g = \Sigma_{g,0}(R/R_0)^{-p}$ and temperature profile $T_g = T_{g,0}(R/R_0)^{-q}$ as a function of the distance to the star $R$, between an inner radius $R_{in}$ and an outer radius $R_{out}$. We assume the gas disc is in steady state, with a vertically isothermal profile.

## 2.2 Radial drift

To take into account dust drift due to gas drag, we use for the radial velocity of dust grains (Dipierro & Laibe 2017; Kanagawa et al. 2017)

$$v_{d,R} = \frac{St}{(1+\epsilon)^2 + St^2} v_{drift} + \frac{1+\epsilon}{(1+\epsilon)^2 + St^2} v_{visc}, \quad (1)$$

valid in the case of a non-self-gravitating and stationary disc surrounding a single star. This expression includes the back-reaction, i.e. the drag of dust on gas, through the dust-to-gas ratio $\epsilon$. St is the Stokes number (see equation 10). The first term is the radial drift velocity with respect to the gas, related to the gas pressure gradient via (Nakagawa, Sekiya & Hayashi 1986):

$$v_{drift} = \left(\frac{H}{R}\right)^2 \frac{d \ln P_g}{d \ln R} v_K, \quad (2)$$

where $\frac{H}{R}$ is the disc aspect ratio, $P_g$ the gas pressure, and $v_K$ the Keplerian orbital velocity. The second term is an additional drag term caused by the radial motion of the gas induced by viscosity, with velocity $v_{visc}$ (Lynden-Bell & Pringle 1974), reformulated for a Keplerian disc:

$$v_{visc} = -3 \frac{\frac{d}{dR}\left(\nu \rho_g R v_K\right)}{\rho_g R v_K}, \quad (3)$$

where $\rho_g$ is the gas density. The gas viscosity $\nu$ is related to the turbulent viscosity parameter $\alpha$ (Shakura & Sunyaev 1973) by $\nu = \alpha H^2 v_K / R$. The second term vanishes if one neglects the gas viscous drift velocity ($v_{visc} = 0$), recovering the inviscid case treated by Nakagawa et al. (1986). Setting $\epsilon = 0$ in equation (1) amounts to neglecting back-reaction, recovering the expression used by, e.g. Birnstiel, Dullemond & Brauer (2010):

$$v_{d,R} = \frac{St}{1+St^2} v_{drift} + \frac{1}{1+St^2} v_{visc}. \quad (4)$$

PACED assumes both $\epsilon = 0$ and $v_{visc} = 0$ (Garcia & Gonzalez 2020).

## 2.3 Dust grain growth model

In this study, we focus on the evolution of porous aggregates. To model their growth, we consider a locally mono-disperse mass distribution where collisions occur between identical grains, as in Laibe et al. (2008). Each collision doubles the mass $m$ of a grain of size $s$ in a mean time $\tau_{coll}$, resulting in the expression given by

Stepinski & Valageas (1997):

$$\left(\frac{dm}{dt}\right)_g \approx \frac{m}{\tau_{coll}} = 4\pi \rho_d s^2 v_{rel}, \quad (5)$$

with $\rho_d$ the local dust density. Grains collide with a relative velocity $v_{rel}$ due to the gas turbulence transmitted to the dust by drag. In this model, relative motions due to friction in radial, azimuthal, and vertical direction vanish by definition, as the friction applied on identical grains at a given location is the same in all directions. Brownian motion caused by thermal agitation is the main source of relative velocity only for very small grains, typically sub-µm grains. As grain growth is extremely fast at these sizes, we neglect this contribution (see Vericel & Gonzalez 2020, for a discussion). $v_{rel}$ can be expressed as (Stepinski & Valageas 1997)

$$v_{rel} = \sqrt{2} v_t \frac{\sqrt{Sc-1}}{Sc}. \quad (6)$$

The turbulent velocity is given by

$$v_t = \sqrt{2^{1/2} Ro \, \alpha} \, c_g \quad (7)$$

with $c_g$ the gas sound speed and Ro, the Rossby number, is considered to be a constant equal to 3. Sc is the Schmidt number of a dust grain, related to the Stokes number by

$$Sc = (1+St)\sqrt{1 + \frac{\Delta v^2}{v_t^2}}, \quad (8)$$

where $\Delta v = v_d - v_g$ is the difference between dust and gas velocities. To express the coupling between gas and dust, one defines the stopping time $\tau_s$, which corresponds to the time needed for a grain to reach the gas velocity:

$$\tau_s = \frac{m \Delta v}{|F_D|}. \quad (9)$$

$F_D$ is the drag force from the gas on dust grains. The Stokes number St is defined as the product of $\tau_s$ and the orbital frequency $\Omega_K$.

$$St = \tau_s \Omega_K. \quad (10)$$

Depending on the value of St, grains behave differently. If St $\ll 1$, dust grains are typically small and well coupled to the gas while St $\gg 1$ corresponds to dust grains large enough to be almost completely decoupled from the gas. However, when St $\approx 1$, dust grains have an intermediate size, are marginally coupled and their radial drift is strongest. Depending on the drag regime, related to the mean free path of the gas $\lambda$: Epstein (Epstein 1924) ($s < 9\lambda/4$) or Stokes (Whipple 1972) ($s > 9\lambda/4$), the Stokes number can be written as:[1]

$$St_{Ep} = \frac{\rho s}{\rho_g c_g} \Omega_K, \quad (11)$$

$$St_{St} = \begin{cases} \dfrac{2\rho s^2}{9\rho_g v_m} \Omega_K, & \text{(linear, Re} < 1) \\[2mm] \dfrac{\rho s}{9\rho_g |\Delta v|} (Re)^{0.6} \Omega_K, & \text{(non-linear, } 1 < Re < 800) \\[2mm] \dfrac{8\rho s}{1.32\rho_g |\Delta v|} \Omega_K. & \text{(quadratic, Re} > 800) \end{cases} \quad (12)$$

---

[1] Some authors (e.g. Ormel & Cuzzi 2007) take the collisional velocities of gas molecules on grains involved in the drag force calculation in the Epstein regime to be equal to the gas sound speed $c_g$, as do we, while others (e.g. Birnstiel et al. 2010) consider it to be the mean gas thermal velocity, which adds a numerical factor $\sqrt{\pi/8}$.

We denote $\rho$ the mean internal density of a grain, $s$ its radius, and $\mathrm{Re} = 2s|\Delta v|/\nu_{\mathrm{m}}$ the Reynolds number (Whipple 1972). Contrary to Garcia & Gonzalez (2020), we consider all three Stokes regimes, not only the linear one. The gas molecular kinematic viscosity $\nu_{\mathrm{m}}$ is defined with the Chapman–Enskog theory and the Sutherland model applied to rigid elastic spheres (Chapman & Cowling 1970; Stepinski & Valageas 1996; Laibe, Gonzalez & Maddison 2012; Wiranata, Prakash & Chakraborty 2012) as

$$\nu_{\mathrm{m}} = \frac{5\sqrt{\pi}}{64} \frac{m_{\mathrm{g}} c_{\mathrm{g}}}{\sigma_{\mathrm{mol}} \rho_{\mathrm{g}}}, \tag{13}$$

with $m_{\mathrm{g}} \approx 3.85 \times 10^{-27}$ kg the mean molecular mass of the gas and $\sigma_{\mathrm{mol}} = 2 \times 10^{-19}$ m$^2$ the cross-section of the H$_2$ molecule. Finally, to compute $\Delta v$, we use

$$\Delta v_r = \frac{(1 + \epsilon)\,\mathrm{St}}{(1 + \epsilon)^2 + \mathrm{St}^2} v_{\mathrm{drift}} - \frac{\mathrm{St}^2}{(1 + \epsilon)^2 + \mathrm{St}^2} v_{\mathrm{visc}}, \tag{14}$$

$$\Delta v_\theta = -\frac{\mathrm{St}^2}{(1 + \epsilon)^2 + \mathrm{St}^2} \frac{v_{\mathrm{drift}}}{2} - \frac{(1 + \epsilon)\,\mathrm{St}}{(1 + \epsilon)^2 + \mathrm{St}^2} \frac{v_{\mathrm{visc}}}{2}, \tag{15}$$

the radial (equation 14) and azimuthal (equation 15) velocity difference between dust and gas (Dipierro & Laibe 2017; Kanagawa et al. 2017). Hence, $\Delta v = \sqrt{\Delta v_r^2 + \Delta v_\theta^2}$. Garcia & Gonzalez (2020) consider in PACED the orbital velocity of dust for all sizes to be $v_{\mathrm{K}}$, leading to an incorrect $\Delta v_{\theta} = v_{\mathrm{drift}}/4$. Tatsuuma & Kataoka (2021) neglect back-reaction, which implies that $\epsilon = 0$, and assume the steady-state minimum mass solar nebula (MMSN) disc. They use simplified versions of equations (14) and (15) with $v_{\mathrm{visc}} = 0$ and $v_{\mathrm{drift}}/2 = 54$ m s$^{-1}$ through the whole disc.

## 2.4 Porosity evolution model

To take into account grain porosity, we use the algorithm derived by Garcia (2018) and Garcia & Gonzalez (2020) based on the models of Suyama et al. (2008), Okuzumi et al. (2009, 2012), and Kataoka et al. (2013).

A dust aggregate is a collection of $n$ monomers considered to be compact spheres of mass $m_0$, size $a_0$, and intrinsic density $\rho_{\mathrm{s}}$. The mass $m$ of the aggregate of size $s$ and mean internal density $\rho$ can be computed as follows:

$$m = \rho V = \rho_{\mathrm{s}} \phi \frac{4\pi}{3} s^3. \tag{16}$$

We suppose here that aggregates are spherical for simplicity, with volume $V$. Taking into account the shape of individual evolving grains is beyond the scope of this paper. By definition, the filling factor $\phi$ is the ratio between the mean internal density and the intrinsic density of the monomers which compose the aggregate.

Two important energies can be associated to the grains. The first one is the kinetic energy when two grains of mass $m$ collide with each other with a relative velocity $v_{\mathrm{rel}}$. In the frame of the centre of mass, the kinetic energy is

$$E_{\mathrm{kin}} = \frac{m}{4} v_{\mathrm{rel}}^2, \tag{17}$$

where the factor 1/4 comes from the reduced mass being half that of the two identical grains. The second one is the rolling energy. It corresponds to the amount of energy necessary to roll a monomer by 90° around a connection point, leading to internal reorganization (Dominik & Tielens 1997):

$$E_{\mathrm{roll}} = 6\pi^2 \gamma_{\mathrm{s}} a_0 \xi_{\mathrm{crit}}, \tag{18}$$

where $\gamma_{\mathrm{s}}$ is the surface energy of a monomer and $\xi_{\mathrm{crit}}$ the critical rolling displacement. Depending on the ratio between the two energies, one can define two different regimes of growth with distinct porosity evolution as a function of the grain's mass $m$. In the 'hit & stick' regime, grains are small and coupled to the gas. This means the collision happens at low relative velocity, with kinetic energy smaller than the rolling energy ($2.2\,E_{\mathrm{kin}} < E_{\mathrm{roll}}$, Suyama et al. 2008). For each collision, the mass doubles and the volume is multiplied by a factor 2/2.99, and the filling factor after an arbitrary and non-integer number of collisions can be expressed as

$$\phi_{\mathrm{h\&s}} = \left(\frac{m}{m_0}\right)^{\ln(2/2.99)/\ln(2)} \tag{19}$$

(Garcia & Gonzalez 2020). As grains grow, the kinetic energy at impact increases, getting larger than the rolling energy. Thus, a certain amount of kinetic energy is dissipated by internal restructuring, leading to compaction. This is the collisional compaction regime. As $v_{\mathrm{rel}}$ depends on the Stokes number, the final filling factor takes a different expression for each drag regime. The equations describing the evolution of porosity in the collisional compression regime and their derivation are presented in Garcia & Gonzalez (2020).

Independently of grain–grain interaction, aggregates can also experience static compaction due either to the gas flow or self-gravity (Kataoka et al. 2013). For the gas flow static compression, one can relate the drag pressure exerted by gas on dust to the filling factor, which leads to

$$\phi_{\mathrm{gas}} = \left(\frac{m_0\,a_0}{\pi E_{\mathrm{roll}}} \frac{\Delta v\,\Omega_{\mathrm{K}}}{\mathrm{St}}\right)^{3/7} \left(\frac{m}{m_0}\right)^{1/7}. \tag{20}$$

For self-gravity, the reasoning is the same, using the relation between the pressure due to self-gravity and the filling factor:

$$\phi_{\mathrm{grav}} = \left(\frac{Gm_0^2}{\pi a_0 E_{\mathrm{roll}}}\right)^{3/5} \left(\frac{m}{m_0}\right)^{2/5}, \tag{21}$$

where G is the universal gravitational constant. Since after its collisional evolution, a grain of mass $m$ can be compacted, its final value of $\phi$ is the maximum of the collisional, gas flow, and self-gravity filling factors.

## 2.5 Fragmentation

When the relative velocity $v_{\mathrm{rel}}$ becomes larger than a threshold $v_{\mathrm{frag}}$, the kinetic energy at impact is sufficient to break bonds between monomers: the grains fragment. To take into account the fragmentation of dust aggregates, Gonzalez et al. (2015) modelled the fragmentation rate in a symmetric way by taking the opposite of the growth rate, i.e. $(\mathrm{d}m/\mathrm{d}t)_{\mathrm{f}} = -(\mathrm{d}m/\mathrm{d}t)_{\mathrm{g}}$. This formulation results in complete fragmentation, most of the mass being lost whatever the value of $v_{\mathrm{rel}}$ compared to $v_{\mathrm{frag}}$. We use here the model developed by Kobayashi & Tanaka (2010) and Garcia (2018), and used by Vericel et al. (2021), where the mass variation is more progressive:

$$\left(\frac{\mathrm{d}m}{\mathrm{d}t}\right)_{\mathrm{f}} = -\frac{v_{\mathrm{rel}}^2}{v_{\mathrm{rel}}^2 + v_{\mathrm{frag}}^2} \left(\frac{\mathrm{d}m}{\mathrm{d}t}\right)_{\mathrm{g}}. \tag{22}$$

This way, fragmentation is more realistic and the grain's mass loss is small close to the threshold, while for large $v_{\mathrm{rel}}$, it is identical to the symmetric model of Gonzalez et al. (2015). After a collision, a fragmenting grain therefore loses half of its mass or more.

We consider here that the filling factor of the aggregate after fragmentation is the same as the initial filling factor, and that the remaining kinetic energy is used to break monomer bonds. According to Sirono (2004), the filling factor remains constant, while Ringl et al. (2012) and Gunkelmann, Ringl & Urbassek (2016) find that the filling factor is multiplied by a factor between one and three. Thus, further studies are needed to take into account grain compaction during fragmentation.

## 2.6 Rotational disruption

Recently, Tatsuuma & Kataoka ([2021](#)) presented a new barrier to dust growth: the rotational disruption barrier. Rotational disruption has already been investigated for interstellar medium dust (Hoang et al. [2019](#)) or cometary dust (Tung & Hoang [2020](#)), but not in the case of protoplanetary discs. We decide to investigate whether disruption occurs before or after the onset of fragmentation, under which circumstances it happens, and whether disruption plays a role in dust growth. Following Tatsuuma & Kataoka ([2021](#)), we suppose that our grains are always in a steady-state angular velocity regime to be able to compute the angular velocity $\omega_c$. We assume that $\omega_c$ is driven only by the gas-flow torque. Indeed, the radiative torque contribution to the total spin period is found to range from two to several orders of magnitude below the gas-flow torque contribution. We also consider a relatively weak turbulent gas, as strong turbulence has unknown effects on whether grains are disrupted or not due to non-trivial gas flows.

To compute $\omega_c$, we use the condition of the steady-state angular velocity $d\omega_c/dt = 0$, i.e. when the spin-up torque due to the gas flow

$$\Gamma_{up} = \frac{2 s F_D \gamma_{ft}}{3} \tag{23}$$

is equal to the spin-down torque caused by collisions with gas particles

$$\Gamma_{down} = -\frac{\omega_c I}{\tau_s}, \tag{24}$$

where

$$I = \frac{8}{15} \pi \rho_s \phi s^5 \tag{25}$$

is the moment of inertia of a spherical dust grain. The term $\gamma_{ft}$ is the force-to-torque efficiency of the gas flow on the aggregate. From equations ([23](#)), ([24](#)), and ([25](#)), the steady-state angular velocity can be derived:

$$\omega_c = \frac{5 \gamma_{ft} \Delta v}{3 s}. \tag{26}$$

We then compute the tensile stress (Hoang et al. [2019](#)):

$$S = \frac{\phi \rho_s s^2 \omega_c^2}{4} \tag{27}$$

and we compare it to the tensile strength

$$S_{max} = 6 \times 10^5 \left(\frac{\gamma_s}{0.1\,\mathrm{J\,m^{-2}}}\right) \left(\frac{a_0}{0.1\,\mu m}\right)^{-1} \phi^{1.8}\,\mathrm{Pa}, \tag{28}$$

derived by Tatsuuma et al. ([2019](#)) to determine whether a grain is rotationally disrupted in our simulations i.e. when $S \gtrsim S_{max}$. Kimura et al. ([2020](#)) also derived an expression to compute the tensile strength of materials which takes into account the aggregate's volume effect:

$$S_{max} = 8 \left(\frac{\gamma_s}{0.1\,\mathrm{J\,m^{-2}}}\right) \left(\frac{a_0}{0.1\,\mu m}\right)^{3/k-1} \left(\frac{\phi}{0.1}\right)^{1.5-1/k} \times$$
$$\exp\left(0.24\left(\frac{\phi}{0.1}-1\right)\right) \left(\frac{V}{686\,\mu m^3}\right)^{-1/k}\,\mathrm{kPa}, \tag{29}$$

with $V$ the volume of a spherical grain. The Weibull modulus $k$ essentially defines the variability of strength inside a material. For silicates $k = 8$ and for water ice, $k = 5$ according to Petrovic ([2003](#)) and Klein ([2009](#)). We assume equations ([28](#)) and ([29](#)) to be valid for the whole range of aggregate's volume we have simulated to be able to study the influence of the tensile strength modelling on the grain evolution. When the tensile stress rises above its tensile strength, the simulation is stopped, as there is currently no model predicting the size of the fragments.

**Table 1.** Disc models used in the simulations.

| Model | $p$ | $q$ | $\Sigma_{g,0}$ (kg m$^{-2}$) | $T_{g,0}$ (K) | $R_{in}$ (au) | $R_{out}$ (au) |
|-------|-----|-----|------------------------------|---------------|---------------|----------------|
| MMSN  | 1.5 | 0.5 | 17000                        | 280           | 1             | 100            |
| Std   | 1   | 0.5 | 488                          | 200           | 1             | 300            |

## 2.7 Set-up

We choose here to use two models: the MMSN model (Hayashi [1981](#)) and a model of disc we call Std (for Standard) that represents an average observed disc from (Williams & Best [2014](#)). Their parameters are given in Table [1](#), where quantities denoted by a 0 are taken at the reference radius $R_0 = 1$ au. The star mass is set to $M_{star} = 1\,M_\odot$, the total mass of the MMSN disc is $M_{disc} = 0.022\,M_\odot$ and that of the Std disc is $M_{disc} = 0.0103\,M_\odot$ (obtained by integrating the disc surface density between $R_{in}$ and $R_{out}$). In this paper, we choose to neglect the back-reaction of dust on gas as the dust-to-gas ratio $\epsilon$ is kept fixed and only one grain evolves at a time. To study the effect of disruption, we choose to investigate the effect of the monomer size $a_0$ with various turbulent viscosity parameters $\alpha$ to compute when grains are disrupted. We choose two different monomer sizes: $a_0 = 0.1$ and 1 $\mu m$ compatible with measurements (Güttler et al. [2019](#); Tazaki & Dominik [2022](#)). $a_0 = 0.1\,\mu m$ is our fiducial value, and we mention the monomer size only when it is needed. We select two relevant species in this study, water ice and silicates. The intrinsic density of water ice monomers is $\rho_s = 1000\,\mathrm{kg\,m^{-3}}$ with a surface energy of $\gamma_s = 0.1\,\mathrm{J\,m^{-2}}$. As the critical rolling displacement is uncertain and still under debate, we choose $\xi_{crit} = 8$ Å (Wada et al. [2011](#); Tatsuuma & Kataoka [2021](#)), and the fragmentation threshold for water ice is set to $v_{frag,H2O} = 15\,\mathrm{m\,s^{-1}}$ (e.g. Gonzalez et al. [2015](#), see also Section [4.4](#)). To model silicate aggregates, we choose an intrinsic density $\rho_s = 2700\,\mathrm{kg\,m^{-3}}$ and a surface energy $\gamma_s = 0.3\,\mathrm{J\,m^{-2}}$ according to Yamamoto, Kadono & Wada ([2014](#)), estimated by a relation between $\gamma_s$ and the melting temperature of the material. This is of the same order of magnitude as the value of $\gamma_s = 0.15\,\mathrm{J\,m^{-2}}$ adopted by Kimura et al. ([2015](#)) using experimental measurements of sicastar aggregates (micromod Partikeltechnologie GmbH). The Young modulus $\mathcal{E}$ is 72 GPa (Yamamoto et al. [2014](#)), which gives, using the assumption that the critical distance $\delta_c$ between two monomers before they separate is of the same order of magnitude as $\xi_{crit}$ (Chokshi, Tielens & Hollenbach [1993](#)), $\xi_{crit} \approx 6$ Å.[2] For silicates, the fragmentation velocity is usually taken as $v_{frag,Si} \sim 1\,\mathrm{m\,s^{-1}}$ (e.g. Birnstiel et al. [2010](#)). However, Yamamoto et al. ([2014](#)) and Kimura et al. ([2015](#), [2020](#)) have shown that silicates are much more resistant than previously thought. We thus choose to take $v_{frag,Si} = 10\,\mathrm{m\,s^{-1}}$ instead of $1\,\mathrm{m\,s^{-1}}$, which is of the same order of magnitude as water ice. We assume the same values of fragmentation thresholds and surface energies of water ice and silicates apply for both monomer sizes. See Section [4.4](#) for a discussion of uncertainties. Turbulence is a key parameter for the relative velocity between grains at impact, hence we took a wide range of turbulence parameters $\alpha = 5 \times 10^{-3}$ to $5 \times 10^{-6}$. Finally, we set the force-to-torque efficiency to $\gamma_{\mathrm{fi}} = 0.1$, its fiducial value from Tatsuuma & Kataoka (2021).

---

[2]This value can be found assuming $\xi_{crit} = \delta_c = \left(\frac{27\pi}{2} \frac{\gamma_s^2 a_0}{\mathcal{E}^2}\right)^{1/3}$.

## 3 RESULTS

### 3.1 The case of the MMSN model

As a first step to study the importance of rotational disruption, we run simulations where the grains are static and allowed to grow from monomers, but not to fragment, up to the disruption barrier. We first compare our results with those obtained by Tatsuuma & Kataoka (2021) in Table 2, showing the properties of dust aggregates when they are rotationally disrupted, at selected fixed locations in the disc and for two monomer sizes. We adopt a very low turbulence $\alpha = 10^{-10}$ to better compare with Tatsuuma & Kataoka (2021), who neglect turbulence. Table 2 shows that our grains are disrupted at slightly smaller masses and sizes, because of smaller filling factors and therefore lower tensile strength on the one hand, and larger $\omega_c$ increasing the tensile stress on the other hand. Differences are likely due to our growth model, which uses the mono-disperse approximation, and to the fact that we compute $v_{\mathrm{drift}}$ every time step, whereas it is a constant in Tatsuuma & Kataoka (2021), slightly increasing $\Delta v$ and thus $\omega_c$. However, the orders of magnitude are in good agreement.

We then run simulations for $\alpha = 5 \times 10^{-6}$, $5 \times 10^{-5}$, $5 \times 10^{-4}$, and $5 \times 10^{-3}$ and for $a_0 = 0.1$ and 1 μm, for grains at fixed locations ranging from $R_{\mathrm{in}}$ to $R_{\mathrm{out}}$. The top panel of Fig. 1 shows the grain size at the disruption barrier as a function of the distance to the star. We find that aggregates made of 1 μm monomers are disrupted earlier in their growth than those made of 0.1 μm monomers. Thus, grains are more affected by rotational disruption when monomer size increases, as shown by Tatsuuma & Kataoka (2021). Four different slopes can be identified for each case, they correspond to the compression and drag regime the grain is in just before the disruption. Following the curve in the top panel of Fig. 1 for $a_0 = 1$ μm and $\alpha = 5 \times 10^{-6}$, grains are, from the inner radius to the outer one, in the gas compression regime for the first three slopes and in the non-linear Stokes, linear Stokes, and Epstein drag regime, respectively, then in the collisional compression regime and Epstein drag regime. This is valid for all cases except $\alpha = 5 \times 10^{-3}$ and $a_0 = 1$ μm where the gas compression regime is never reached between 1 and 10 au (see the blue dashed line of Fig. 1). Instead, grains are found to disrupt in the collisional compression regime.

Turbulence plays a role only in the outer part of the disc, i.e. in the collisional compression regime, because gas compression does not depend on the turbulence $\alpha$ though $\Delta v$. [To be precise, $|v_{\mathrm{visc}}/v_{\mathrm{drift}}| \sim \alpha$ – see appendix C of Gonzalez et al. (2017). The contributions of $v_{\mathrm{visc}}$ relative to those of $v_{\mathrm{drift}}$ in equations (14) and (15) are thus of order $\alpha$St and $\alpha/$St, respectively, and negligible for the range of St values ($\sim 0.1$–1) corresponding to the maximum sizes reached in the simulations.] Increasing the turbulence leads to disruption earlier in the grain's growth, truncating the area where aggregates can grow up to meter size and above. This is due to the fact that a higher $\alpha$, and thus a higher $v_{\mathrm{rel}}$ (equations 6 and 7) leads to a more efficient collisional compression. At large radii, gas flow compression is replaced by collisional compression as the main source of compaction before rotational disruption. In the inner disc, as the gas compression regime does not depend on turbulence, disruption is the same whatever the turbulence. However, it is still dependent on the monomer size as it directly affects the compactness of grains and its ability to resist to gas compaction.

The bottom panel of Fig. 1 shows the relative velocity between grains when they are disrupted. This allows to determine when grains are shattered by collision or disruption. The typical fragmentation threshold of water ice is plotted for reference. For a high turbulence of $\alpha = 5 \times 10^{-3}$, we can deduce that grains will fragment by collisions before being rotationally disrupted while for $\alpha \leq 5 \times 10^{-5}$, only the disruption prevails. In between, grains are destroyed by collisional fragmentation in the first few (5–10) au and by rotational disruption in the rest of the disc. Interestingly, the monomer size does not have a strong influence on the relative velocity at which grains are disrupted. Again, the change in compression regime is responsible for the slope change between different disc regions. For a very low or null viscosity, rotational disruption is the dominant mechanism to destroy grains. When $\alpha$ reaches values on the order of $10^{-2}$–$10^{-3}$, this is not true anymore (consistent with equation 36 of Tatsuuma & Kataoka 2021) and depending on the position in the disc and on the viscosity, the outcome can be different.

Until now, we did not take into account the grain's spatial evolution in the disc as its orbital position was kept fixed. Fig. 2 now shows the evolution of grains at multiple initial locations within the disc, taking into account radial drift. In these simulations, aggregates are able to grow, fragment, and drift radially; however, we did not include rotational disruption in their evolution. The objective is to see under which conditions grains can grow above the disruption limit, and whether neglecting disruption changes the evolution of dust radically. Conforming to the previous result, grains in a disc with $\alpha = 5 \times 10^{-3}$ are shattered by collisional fragmentation only. The grains start to grow from the monomer size, then slowly drift as St increases, causing $v_{\mathrm{rel}}$ to increase and rapidly exceed the fragmentation threshold. An equilibrium between growth and fragmentation is reached. As aggregates fragment at sizes for which the Stokes number is larger than $10^{-2}$, radial drift is fast and all the grains are accreted on the star without reaching the disruption limit. Lowering the viscosity parameter to $\alpha = 5 \times 10^{-4}$ still allows collisional fragmentation, but only in the inner part of the disc, between 1 and 6–7 au. Past these radii, grains are shattered by rotational disruption, as the relative velocity is not high enough to reach collisional fragmentation. In the case of $\alpha = 5 \times 10^{-5}$, the turbulent viscosity and therefore the relative velocity between grains is too small to allow collisional fragmentation before accretion. Thus, dust grains grow at fixed locations up to St $\approx 10^{-2}$, then drift radially until they cross the disruption limit where they should be destroyed (the dashed lines in Fig. 2 show their subsequent evolution above the disruption limit when it is not included). We observe the same evolution pattern with a monomer size of $a_0 = 1$ μm (see Fig. A1). In the case of the MMSN disc model, grains are destroyed by collisional fragmentation for $\alpha \geq 5 \times 10^{-3}$ and by rotational disruption below.

### 3.2 Effect of grain material in the Std disc

We consider in the rest of this paper our Std disc model (see Table 1), which fits disc observations better. We investigate here the effect of rotational disruption for two different species: water ice like in Section 3.1 and silicates, another very common material in discs, to understand if different material properties change significantly the limit between disruption and fragmentation.

Similarly to Fig. 1, Fig. 3 shows the disruption barrier in grain size (top panel) and relative velocity between water ice aggregates (bottom panel) as a function of the distance to the star. We observe the same tendency in this disc model compared to MMSN, i.e. a higher turbulence leads to rotational disruption at smaller sizes compared to lower turbulence, for which sizes of tens of meters can be reached in a much wider region for aggregates with 0.1 $\mu$m monomers. We can see the same slope change as in Fig. 1 due to the different transitions between collisional and gas compression regimes, but shifted toward smaller radii. One should note that gas compression in the non-linear Stokes regime does not appear for $a_0 = 1$ $\mu$m whatever $\alpha$, and that grains are in the collisional compression regime, not only for $\alpha = 5 \times 10^{-3}$ as found also in Fig. 1, but also for $\alpha = 5 \times 10^{-4}$. Likewise, the relative velocity does not depend strongly on the monomer size $a_0$.

**Table 2.** Comparison of aggregate properties when they are rotationally disrupted between Tatsuuma & Kataoka (2021) and this work.

| $R$ (au) | $a_0$ ($\mu$m) | $m$ (kg) | $\phi$ | $s$ (m) | St | $\omega_c$ (rad s$^{-1}$) | $S$ (Pa) |
|---|---|---|---|---|---|---|---|
| | | | Tatsuuma & Kataoka (2021) | | | | |
| 5 | 0.1 | $4 \times 10^4$ | $4 \times 10^{-4}$ | 30 | 0.1 | $7 \times 10^{-2}$ | 0.5 |
| 10 | 0.1 | $1 \times 10^5$ | $3 \times 10^{-4}$ | 50 | 0.1 | $4 \times 10^{-2}$ | 0.3 |
| 10 | 1 | $2 \times 10^4$ | $2 \times 10^{-3}$ | 10 | 0.07 | $9 \times 10^{-2}$ | 0.6 |
| 50 | 0.1 | $6 \times 10^3$ | $7 \times 10^{-5}$ | 30 | 0.06 | $4 \times 10^{-2}$ | 0.02 |
| | | | This work | | | | |
| 5 | 0.1 | $6 \times 10^3$ | $3 \times 10^{-4}$ | 17 | 0.1 | $1.1 \times 10^{-1}$ | 0.29 |
| 10 | 0.1 | $1.1 \times 10^4$ | $2 \times 10^{-4}$ | 23 | 0.1 | $7.5 \times 10^{-2}$ | 0.16 |
| 10 | 1 | $2.1 \times 10^3$ | $1.2 \times 10^{-3}$ | 7 | 0.06 | $1.4 \times 10^{-1}$ | 0.35 |
| 50 | 0.1 | $1.6 \times 10^3$ | $5.7 \times 10^{-5}$ | 19 | 0.06 | $5 \times 10^{-2}$ | 0.015 |

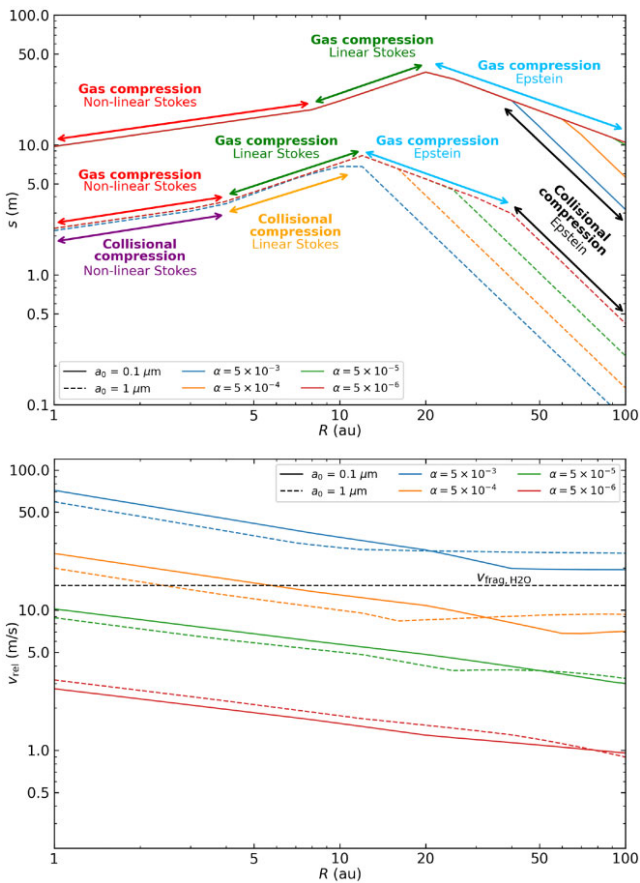

**Figure 1.** Maximum grain size (top panel) and maximum relative velocity between dust grains (bottom panel) before disruption for different $\alpha$ viscosity parameters and monomer sizes $a_0$ with static water ice grains for the MMSN disc. The black dashed line corresponds to the typical fragmentation threshold of water ice.

Due to a higher intrinsic density, silicate grains are more compact for a given mass than water ice grains, as $\phi$ depends on $\rho_s$, which implies smaller sizes. Despite that, the disruption limits are quite similar to that of water ice, even if we note that silicate aggregates tend to be disrupted at sizes smaller by a factor two to three compared to water ice grains (see Fig. A2).

If we now take into account the dust spatial evolution, grain material has a significant impact. Figs 4 and 5 show the evolution of water ice and silicate grains, respectively, at multiple initial locations within the disc taking into account radial drift. For water ice (Fig. 4), we observe the same behaviour as in Fig. 2. The disruption barrier prevails on the fragmentation and radial drift ones for $\alpha \leq 5 \times 10^{-5}$. Grains radially drift until they cross the disruption barrier and are destroyed. The fragmentation and disruption barriers are in competition for $\alpha = 5 \times 10^{-4}$, where grains are destroyed: by fragmentation if they are close to the star ($R \leq 5$ au), and by disruption otherwise. The maximum size grains are able to reach in this case is divided by a factor of two or less if disruption is taken into account. On the other hand, the disruption barrier becomes inefficient comparatively to the fragmentation barrier when turbulence is as high as $\alpha = 5 \times 10^{-3}$. In the case of silicates (Fig. 5), fragmentation is always more efficient, when it occurs, than disruption, even if grains are very close to be disrupted near 15–20 au for $\alpha = 5 \times 10^{-4}$. When lowering the turbulence, the fragmentation barrier disappears in favour of the disruption one. For $\alpha = 5 \times 10^{-5}$, we recover the situation where grains drift up to disruption. As silicate aggregates have a higher intrinsic density, they also drift sooner and faster than their water ice counterparts due to larger St, which explains why grains reached the disruption barrier with more difficulty. The evolution of aggregates of water ice (Fig. A3) or silicates (Fig. A4) made of larger monomers ($a_0 = 1$ $\mu$m) is not different. For water ice and $\alpha = 5 \times 10^{-4}$, the region where disruption operates, is narrower. In the inner region, the fragmentation and the disruption barriers are very close, meaning both barriers can operate in the same location at the same time. Identically to Fig. 4, neglecting disruption does change the maximum size by a factor two to three. For silicates and water ice and $\alpha = 5 \times 10^{-5}$, the maximum grains size is of the same order of magnitude with or without the disruption barrier, contrary to cases with $a_0 = 0.1$ $\mu$m.

### 3.3 Accounting for the aggregate's volume effect

We investigate here the effect of changing the expression of the tensile strength from equation (28) to that derived by Kimura et al.

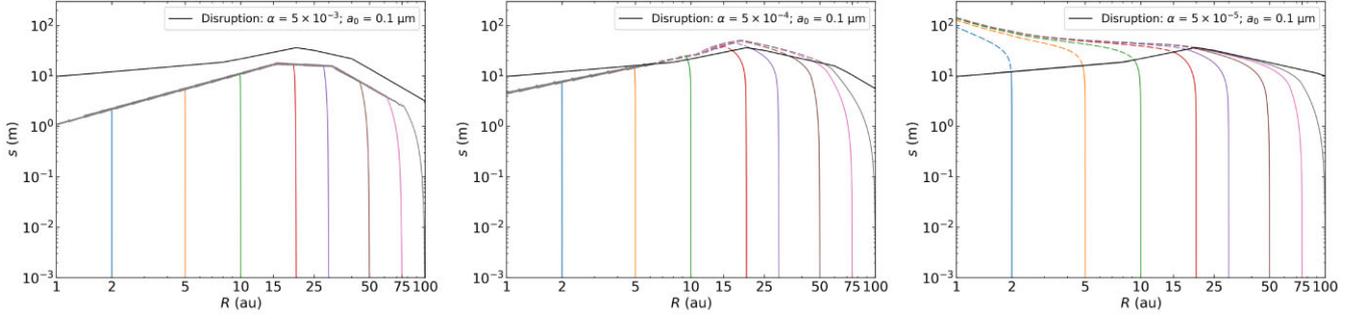

**Figure 2.** Radial and size evolution of water ice grains at multiple initial locations for three viscosity parameters $\alpha = 5 \times 10^{-3}$, $5 \times 10^{-4}$, and $5 \times 10^{-5}$ (from left to right) and $a_0 = 0.1$ μm in the MMSN model. The black line is the disruption limit. The dashed lines show the evolution above the disruption limit when it is not included.

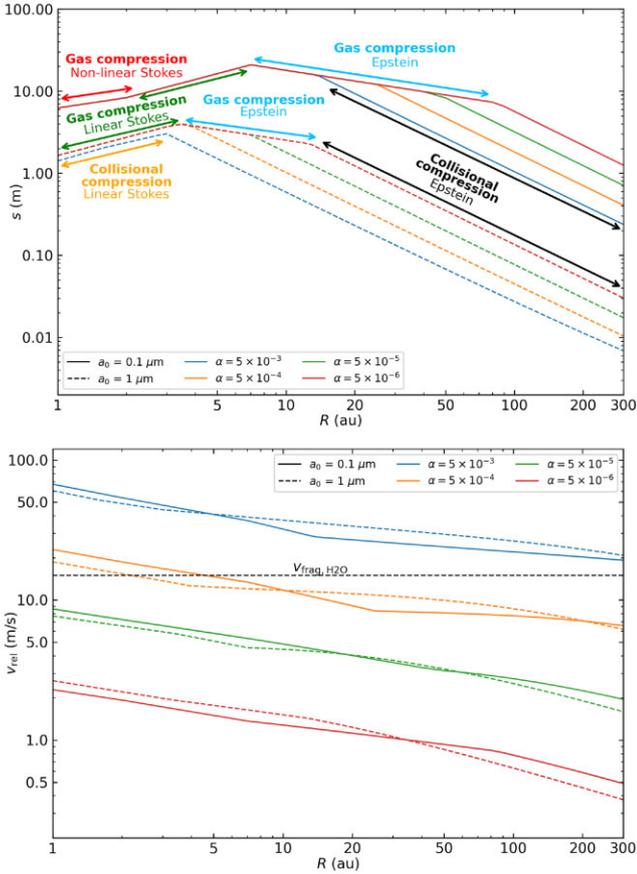

**Figure 3.** Same as Fig. 1 for the Std disc and water ice grains.

(2020) in equation (29). Given the fact the tensile strength is volume-dependent with this equation, we expect easier disruption of grains as they grow, since $S_{max} \propto V^{1/k}$ with $k > 1$. To test this, we use again the Std disc with water ice and silicates. All the other parameters are the same. Tatsuuma & Kataoka (2021) used other values for $\gamma_{ft}$ (a free parameter) which influence the tensile stress $S$ (equation 27). Tatsuuma & Kataoka (2021) estimated, thanks to equation 34 of Lazarian & Hoang (2007) a force-to-torque efficiency $\gamma_{ft} = 0.15$ consistent with the fiducial value. However, lower value of the force-to-torque efficiency are also plausible; thus, we study the disruption limit for both species for three different cases: $\gamma_{ft} = 0.1$, 0.05, and 0.01.

In the case of equation (28), for lower $\gamma_{ft}$ such as 0.05 or 0.01, our simulations reveal that the disruption barrier is simply non-existent, whatever the turbulence or the monomer size (not shown). It is not the case with equation (29), as we will see now. Fig. 6 is similar to Fig. 4; however, this time different disruption limits are plotted. The first three are the limits computed using equation (29) with $\gamma_{ft} = 0.1$, 0.05, and 0.01, and the last one corresponds to equation (28) and $\gamma_{ft} = 0.1$ for reference.

With equation (29), for the fiducial value of $\gamma_{ft} = 0.1$, water ice grains are disrupted whatever the turbulence $\alpha$, before they are able to drift, except for $\alpha \geq 10^{-3}$ between 1 and 3 au where fragmentation dominates. In the case of silicates (Fig. 7), the fragmentation barrier dominates for $\alpha = 5 \times 10^{-3}$, while disruption prevails for lower turbulence. Therefore, if $\gamma_{ft} = 0.1$, disruption is very restrictive, severely inhibiting the growth of aggregates, and the rotational disruption barrier dominates.

If $\gamma_{ft} = 0.05$, the situation is the same for water ice. The disruption barrier prevails for all cases, even for $\alpha = 5 \times 10^{-3}$, except between 1 and 5 au. The fragmentation and disruption barriers become mixed up between 5 and 10 au, i.e. grains would be destroyed by either collisional fragmentation or disruption. However, for silicates, the fragmentation barrier dominates for $\alpha = 5 \times 10^{-3}$, collisional fragmentation being more efficient than for water ice. For $\alpha = 5 \times 10^{-4}$, we observe the same behaviour seen in Fig. 6 for $\alpha = 5 \times 10^{-3}$, where collisional fragmentation dominates in the inner region (1–3 au), while rotational disruption prevails past 10 au. In between, both barriers coexist. $\alpha$ values lower than $5 \times 10^{-4}$ give exactly the same fate encountered in the case of water ice, grains are destroyed by disruption during their radial drift.

If the force-to-torque efficiency is even lower, down to $\gamma_{ft} = 0.01$, disruption is still the barrier that prevents growth of water ice aggregates for $\alpha \leq 5 \times 10^{-4}$. In fact, the situation is the same as the one depicted in Section 3.2, for both water ice and silicates.

In the case of $a_0 = 1$ μm monomers for water ice (see Fig. A5), the situation is slightly different. Collisional fragmentation always destroys grains for $\alpha = 5 \times 10^{-3}$, even with the largest $\gamma_{ft}$. Yet, disruption is still a barrier for all $\gamma_{ft}$ and lower $\alpha$. The case with silicates and $a_0 = 1$ μm in Fig. A6 is quite similar to one with $a_0 = 0.1$ μm. However, we notice one difference between Figs A6 and 7: for $\alpha = 5 \times 10^{-4}$, grains are no longer disrupted when $\gamma_{ft} = 0.05$.

It should be noted that the disruption limit computed using equation (28) and $\gamma_{ft} = 0.1$ is similar to the one computed with equation (29) and $\gamma_{ft} = 0.01$ for water ice, and $\gamma_{ft} = 0.05$ for silicates. The fates of grains for all cases are summarized in Fig. 8.

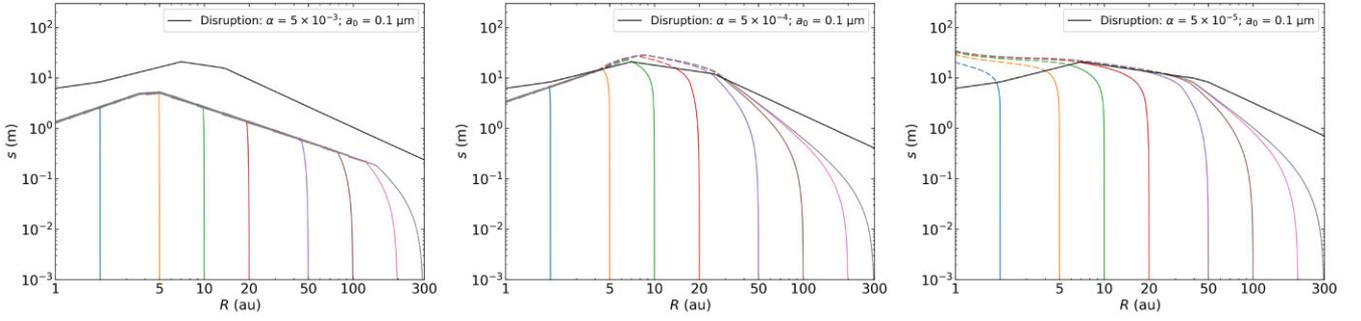

**Figure 4.** Same as Fig. 2 for the Std disc model and water ice grains.

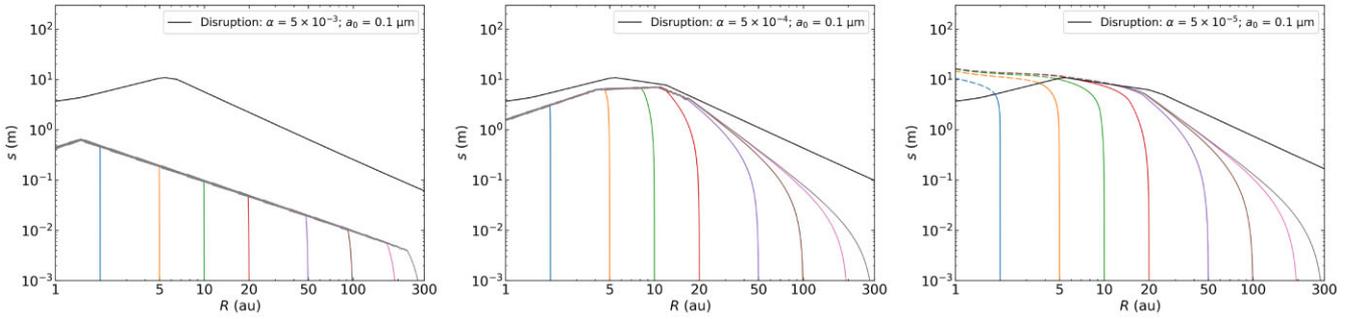

**Figure 5.** Same as Fig. 4 for the Std disc model and silicates grains.

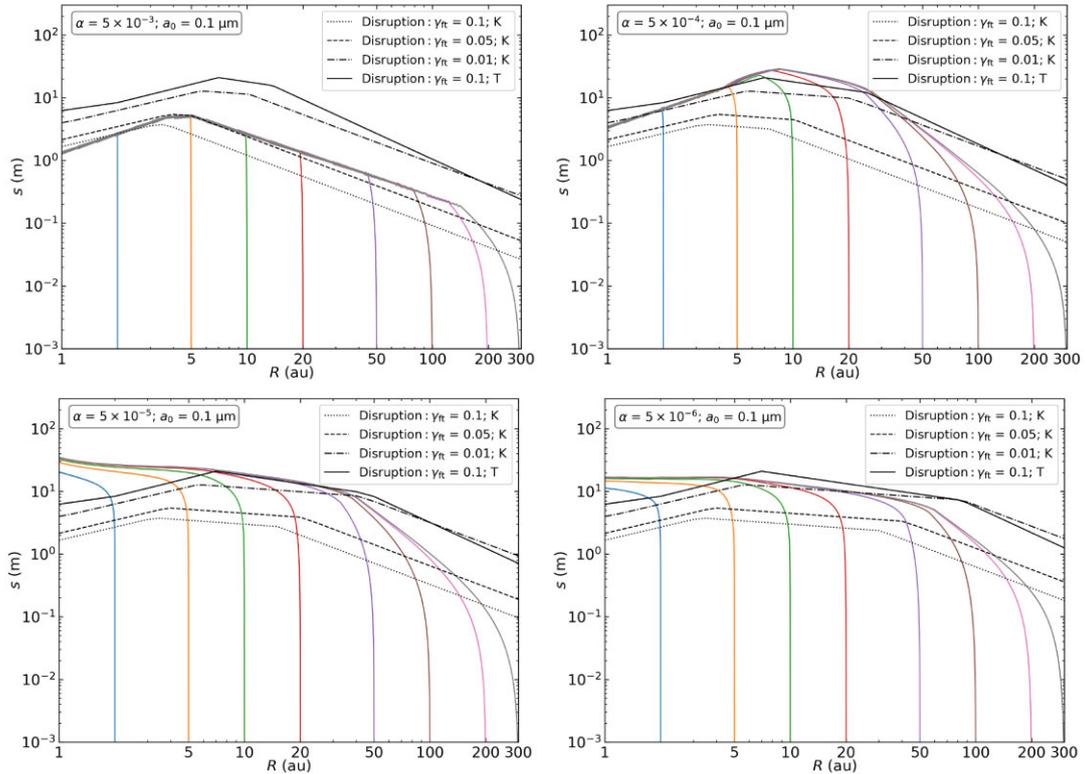

**Figure 6.** Same as Fig. 4 for $\alpha = 5 \times 10^{-3}$, $5 \times 10^{-4}$, $5 \times 10^{-5}$, and $5 \times 10^{-6}$ (from left to right and top to bottom) for the Std disc model and water ice grains. The black lines are the disruption limits where 'K' stands for Kimura et al. (2020) and equation (29) and 'T' for Tatsuuma & Kataoka (2021) and equation (28).

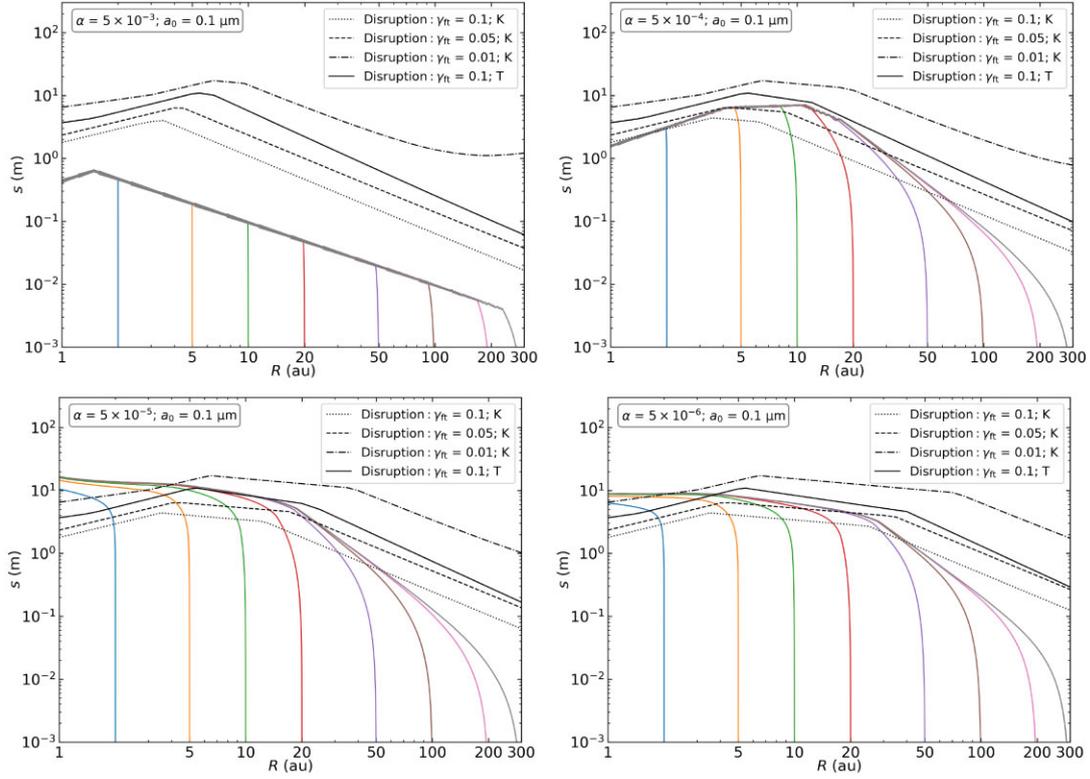

**Figure 7.** Same as Fig. 6 for the Std disc model and silicate grains.

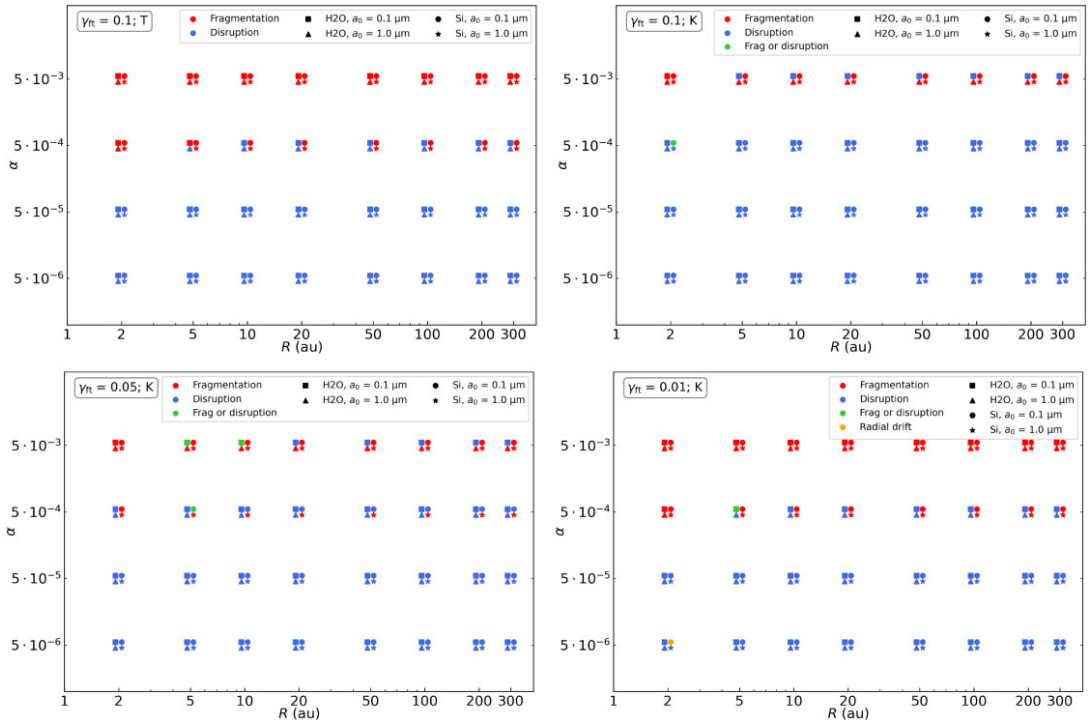

**Figure 8.** Fate of aggregates depending on their initial location and disc turbulent viscosity for all cases with Std discs, with $\gamma_{\rm ft} = 0.1$, 0.05, and 0.01, and where 'K' stands for Kimura et al. (2020) and equation (29) and 'T' for Tatsuuma & Kataoka (2021) and equation (28).

## 4 DISCUSSION

### 4.1 Caveats

Our study has some limitations that arise from our model being 1D in the radial dimension of discs. We also simulate simple power-law discs, neglecting the gas evolution, and grains evolve independently. 3D simulations of discs taking into account both the gas and porous dust evolution may be necessary to understand more precisely how and where disruption influences dust evolution.

Differences between the results of Tatsuuma & Kataoka (2021) and ours are most likely due to the way we model porosity and grain growth. Suyama et al. (2008), Okuzumi et al. (2009), Kataoka et al. (2013), and Tatsuuma & Kataoka (2021) solve the Smoluchowski equation (Smoluchowski 1916). Our code uses the mono-disperse approximation, as it is also designed to straight-forwardly interpret results from 3D simulations of protoplanetary discs with porous grains performed with codes using the same approximation, in future work. Contrary to Kataoka et al. (2013), Tatsuuma & Kataoka (2021) do not consider collisional compaction, as it would require computing gas quantity around their aggregates. We reintroduced grains compaction during their growth depending on the gas-drag regime and Stokes number St, via equations (1), (11), and (12) and the algorithms in appendix A of Garcia & Gonzalez (2020).

Finally, there is to date no study of the number of fragments that are left from a disrupted grain and of their sizes distribution, which would allow us to fully model disruption in global simulations.

### 4.2 Influence of the tensile strength formulation and the force-to-torque efficiency

We show that taking into account the aggregate's volume effect on the tensile strength has a huge effect on the grain evolution for a given $\gamma_{ft}$. Compared to Tatsuuma & Kataoka (2021), the formulation of Kimura et al. (2020) restricts grains to lower sizes at low $\alpha$, where collisional fragmentation is ineffective. Rotational disruption dominates with both equations (28) and (29) which means grains are destroyed even in low-turbulence discs. While the former destroys grains only in the inner regions (where coloured lines cross the black solid line), the latter destroys them in the whole disc (where they cross the black dotted line, see the lower panels in Figs 6, 7, A5, and A6). This radically changes the way grains evolve in discs.

The force-to-torque efficiency is also important as it controls how fast grains are rotationally disrupted during their growth. Tatsuuma & Kataoka (2021) briefly explore this parameter in their simulations, and found that grains are safe from rotational disruption for low values such as $\gamma_{ft} \leq 0.02$. As shown in Section 3.3, lower $\gamma_{ft}$ implies higher disruption sizes, as gas imprints less angular momentum to aggregates.

### 4.3 Is rotational disruption negligible?

We showed thanks to our simulations that grains can be rotationally disrupted depending on the turbulence and monomer size. For the Std or MMSN models, when grains are allowed to move within the disc, the disruption barrier is present mostly for $\alpha \leq 5 \times 10^{-4} - 5 \times 10^{-5}$ for both silicates and water ice. For higher turbulence, collisional fragmentation destroys aggregates, while for lower $\alpha$, it is the radial drift combined with the disruption barrier that prevents grain growth. Therefore, rotational disruption can be neglected for $\alpha \geq 5 \times 10^{-4}$, as the size reached by grains with or without disruption is qualitatively the same. However, for lower turbulence, the fragmentation barrier vanishes as $v_{rel}$ is too small even for large grains; thus, rotational disruption strongly influences the aggregates evolution in the disc's inner region. We show also that, using Kimura et al. (2020) tensile strength formulation, disruption cannot be negligible for high force-to-torque efficiency, especially for water ice grains. However, low $\gamma_{ft}$ renders rotational disruption irrelevant for silicates. Whether the disruption barrier needs to be taken into account depends also on the fragmentation threshold. We choose in this paper a fixed $v_{frag}$; however, other values can also be relevant. A high fragmentation threshold allows grains to grow almost freely, and so rotational disruption will be the main process of destruction. Conversely, a very low threshold will inhibit growth to lower sizes, and operate also at lower turbulence, making the rotational disruption barrier negligible.

### 4.4 The uncertainties of fragmentation thresholds

The fragmentation threshold of disc materials is still an active research field, with many disparities in the establishment of values. Blum & Wurm (2008) and Güttler et al. (2010) give $v_{frag, Si} \sim 1$ m s$^{-1}$. Wada et al. (2009, 2013) found a value close to $v_{frag, Si} \sim 5$ m s$^{-1}$, and $v_{frag, H2O} \sim 60$–70 m s$^{-1}$. Yamamoto et al. (2014) found similar values for ice $v_{frag, H2O} = 56$ m s$^{-1}$, but vastly different ones for silicates $v_{frag, Si} = 55$ m s$^{-1}$. However, the value for the surface energy used by Wada et al. (2009, 2013) is 25 mJ m$^{-2}$, an order of magnitude lower than that found by Yamamoto et al. (2014) and Kimura et al. (2020). This is caused by absorption of water molecules by silicates, which lowers the surface energy. Thus, as the fragmentation velocity $v_{frag} \propto \gamma_s^{5/6}$ (Dominik & Tielens 1997), we choose the value of $v_{frag, Si} = 10$ m s$^{-1}$, of the same order of magnitude as that we adopted for water ice, $v_{frag, H2O} = 15$ m s$^{-1}$. This value was computed by Gonzalez et al. (2015) and Garcia (2018) relying on experimental measurements of fragmentation energy per unit mass ($\sim$55 J kg$^{-1}$) carried out by Shimaki & Arakawa (2012), instead of using values from numerical simulations (Wada et al. 2009).

## 5 CONCLUSIONS

We investigate dust grain shattering in protoplanetary discs to understand if rotational disruption is an important process in aggregate evolution. We wrote a 1D code including our growth + fragmentation and porosity evolution model, as well as radial drift. We incorporate the rotational disruption with two different formulations for the tensile strength. Our code gives results in agreement with Tatsuuma & Kataoka (2021) despite slightly earlier disruption. We then showed that disruption is in competition with collisional fragmentation as soon as the viscosity is lower or equal than $\alpha = 5 \times 10^{-4}$ for the MMSN disc. In the case of the Std disc, similar behaviours are found, for both species and both monomer sizes: For higher turbulence, collisional fragmentation dominates, while for low turbulence, rotational disruption prevails. For intermediate viscosity: $\alpha = 5 \times 10^{-4}$, the destiny of water ice aggregates is to be rotationally disrupted, while it is collisional fragmentation for silicates. However, using the tensile strength formulation of Kimura et al. (2020) depicts a different story. For $\gamma_{ft} = 0.1$, the rotational disruption barrier dictates the evolution of water ice grains for all explored turbulence $\alpha$, while it is not the case for silicates and larger monomers if $\alpha = 5 \times 10^{-3}$. We show that for values $\gamma_{ft} = 0.05$ and $\gamma_{ft} = 0.01$, dust growth is still hampered by fragmentation and disruption.

Nevertheless, further investigation has to be done, mainly to lift some limitations of our code. 3D, two-phase hydrodynamical simulations will allow us to study more precisely where in the disc aggregates are rotationally disrupted, and how they behave in structures like self-induced dust traps or snow-lines when collisional fragmentation also operates. As rotational disruption is of a very different nature than the other two barriers (fragmentation and radial drift), its behaviour in such traps might be unexpected and will be the subject of a future paper.

## ACKNOWLEDGEMENTS

We thank the anonymous referee for their detailed report. The authors acknowledge funding from ANR (Agence Nationale de la Recherche) of France under contract number ANR-16-CE31-0013 (Planet-Forming-Disks) and thank the LABEX Lyon Institute of Origins (ANR-10-LABX-0066) for its financial support within the programme 'Investissements d'Avenir' (ANR-11-IDEX-0007) of the French government operated by the ANR. This project has received funding from the European Union's Horizon 2020 research and innovation programme under the Marie Skłodowska-Curie Actions grant agreements No. 210021 and No. 823823 (DUSTBUSTERS). Figures were made with the Python library `matplotlib` (Hunter 2007).

## DATA AVAILABILITY

PAMDEAS is available on github at this url: https://github.com/Steph aneMichoulier/Pamdeas. The data used for this article will be shared on reasonable request to the corresponding author.

## APPENDIX A: ADDITIONAL FIGURES

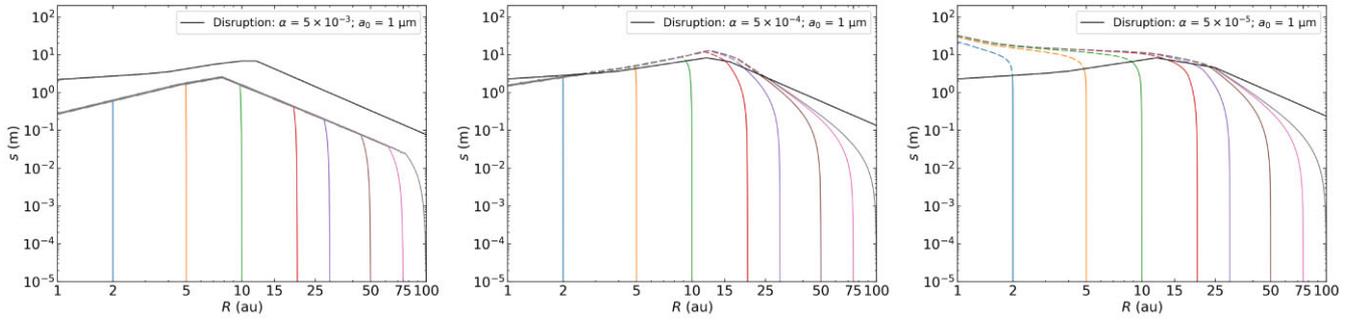

**Figure A1.** Same as Fig. 2 for the MMSN disc model, water ice grains, and $a_0 = 1$ μm. Note the size axis is changed to better fit the data.

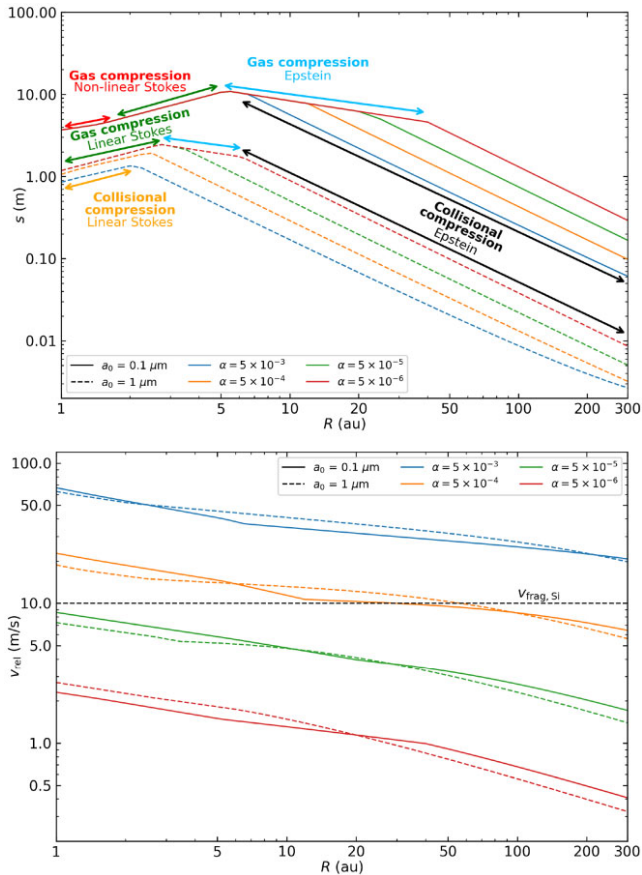

**Figure A2.** Same as Fig. 3 for the Std disc model and silicate grains.

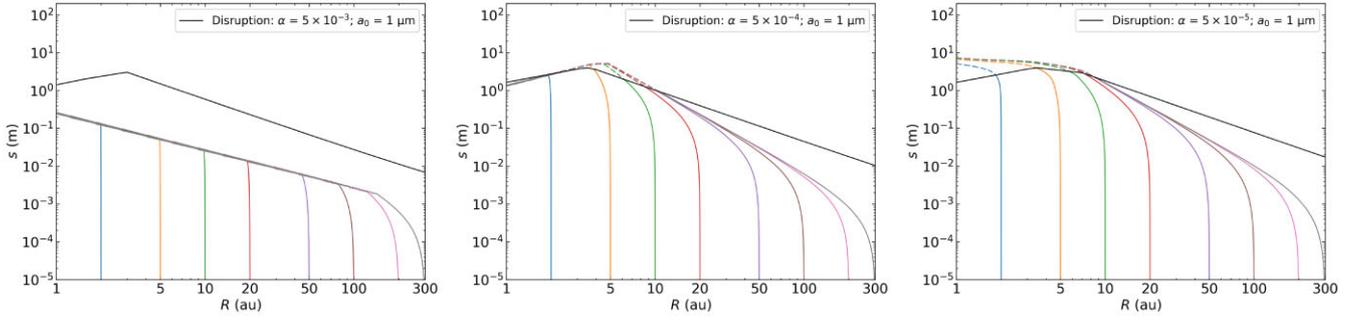

**Figure A3.** Same as Fig. 4 for the Std disc model, water ice grains, and $a_0 = 1\ \mu m$ with different size axis.

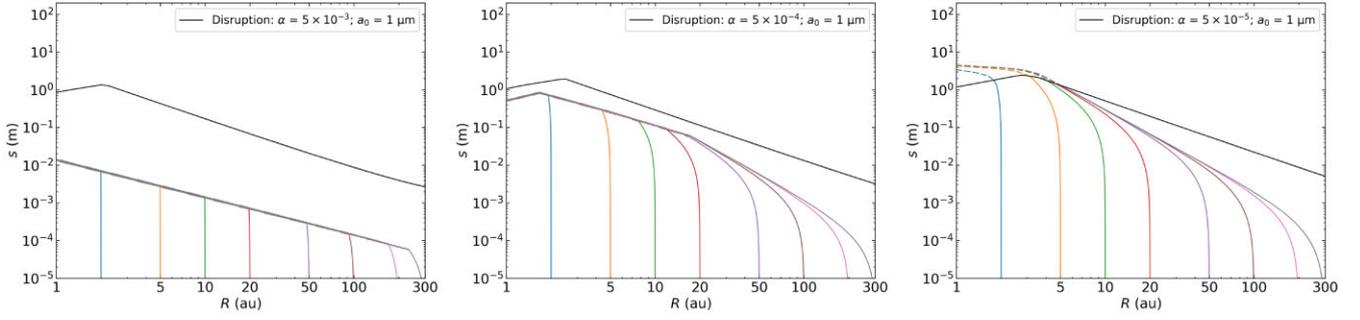

**Figure A4.** Same as Fig. 5 for the Std disc model, silicate grains, and $a_0 = 1\ \mu m$ with different size axis.

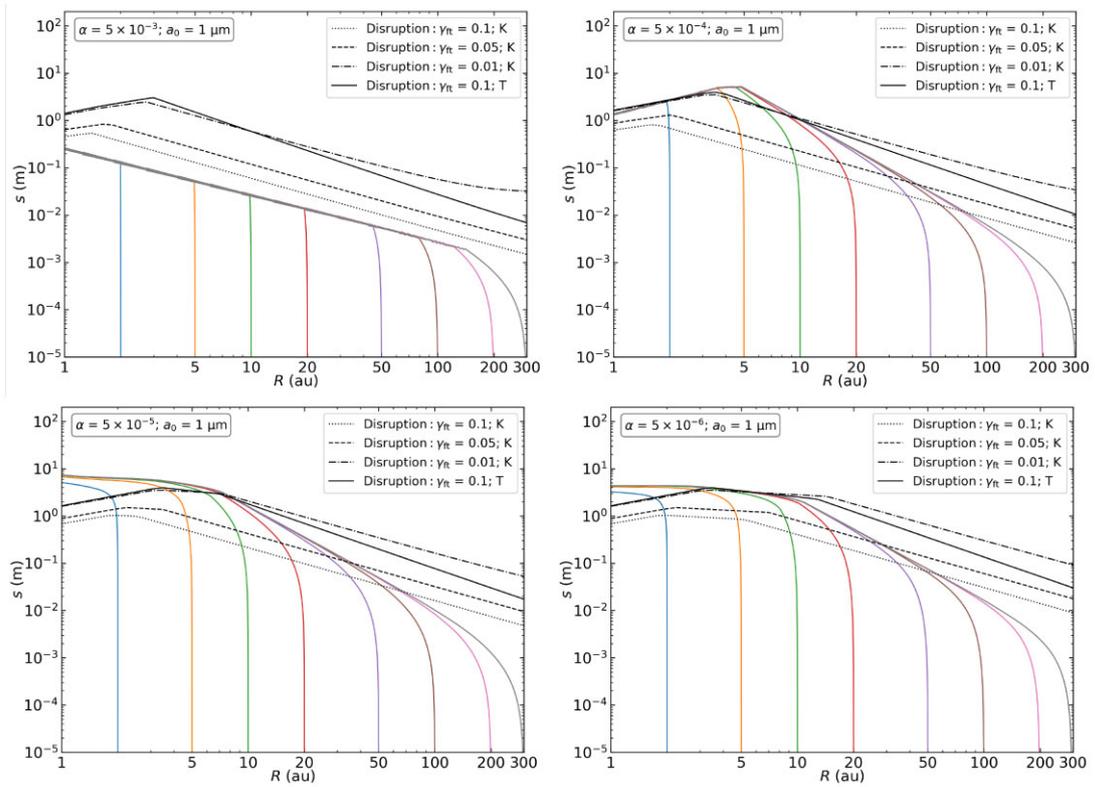

**Figure A5.** Same as Fig. 6 for the Std disc model, water ice grains, and $a_0 = 1\ \mu m$ with different size axis.

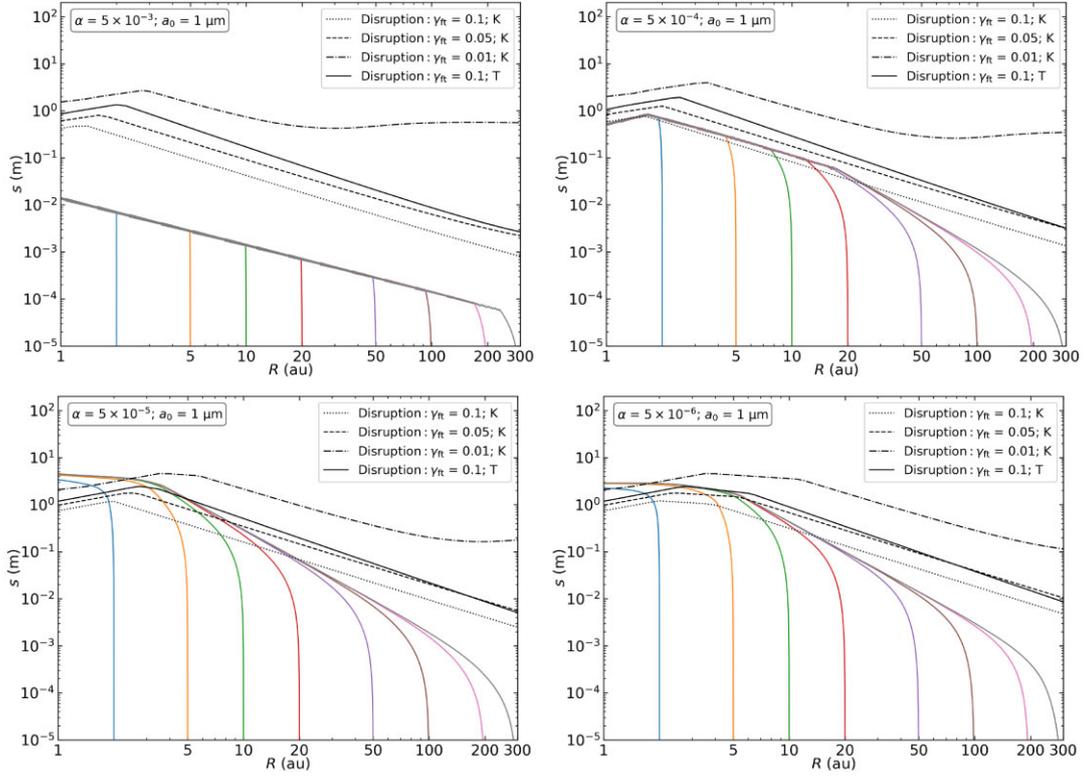

**Figure A6.** Same as Fig. 7 for the Std disc model, silicate grains, and $a_0 = 1$ μm with different size axis.

This paper has been typeset from a TeX/LaTeX file prepared by the author.